\documentclass[useAMS,usenatbib]{mn2e}
\usepackage{epsfig}
\usepackage[totalwidth=480pt, totalheight=680pt]{geometry}

\def\beq{\begin{equation}}
\def\eeq{\end{equation}}
\def\erfi{\mathrm{erfi}}
\def\erf{\mathrm{erf}}
\def\media#1{\langle #1\rangle}

\def\msol{\,\mathrm{M}_\odot}

\title[Energy equipartition in globular clusters models]
{Central energy equipartition in
multi-mass models of globular clusters}
\author[P. Miocchi]{P. Miocchi\thanks{E-mail:
miocchi@uniroma1.it}\\
Dipartimento di Fisica, Universit\'a di Roma ``La Sapienza",\\
P.le Aldo Moro, 2, Rome I00185, Italy.}
\begin{document}

\date{Accepted ????. Received ????; in original form ????}

\pagerange{\pageref{firstpage}--\pageref{lastpage}} \pubyear{????}

\maketitle

\label{firstpage}

\begin{abstract}
In the construction of multi-mass King-Michie models of globular clusters,
an approximated central energy equipartition between stars of different
masses is usually imposed by
scaling the velocity parameter of each mass class inversely with
the stellar mass, as if the distribution function were isothermal.
In this paper, this `isothermal approximation'
(IA) has been checked and its consequences on the model parameters
studied by a comparison with models including
central energy equipartition correctly.
It is found that, under the IA, the `temperatures' of a pair of components
can differ to a non-negligible amount for low concentration distributions.
It is also found that, in general, this approximation leads to a
significantly reduced mass segregation in comparison with that given
under the exact energy equipartition at the centre.
As a representative example, an isotropic
3-component model fitting a given projected surface brightness and
line-of-sight velocity dispersion profiles
is discussed. In this example, the IA gives a cluster envelope much
more concentrated (central
dimensionless potential $W=3.3$) than under the true equipartition
($W=5.9\times 10^{-2}$), as well as a higher logarithmic mass function
slope. As a consequence, the
inferred total mass (and then the global mass-to-light ratio) results
a factor $1.4$ times lower than the correct value and the amount of mass
in heavy dark remnants is $3.3$ times smaller. Under energy
equipartition, the fate of stars having a mass below a certain limit
is to escape from the system. This limit is derived as a function of the
mass and $W$ of the giants and turn-off stars component.
\end{abstract}


\section{Introduction}

Since the end of 60's, models based on the Maxwellian phase-space
distribution function (DF) have been extensively used in the so called
{\it parametric\/} fit of globular clusters light profiles, in order to infer
important properties like the mass-to-light ratio and
the mass function slope \citep[see][sect. 7.7, for a review]{heggiemeylan}.
Corresponding to the isothermal distribution \citep{BT}, the
Maxwellian is expected to arise from relaxation phenomena like
the interaction between the stars and the rapidly varying mean potential
during the `violent relaxation' subsequent to the cluster formation
\citep{lynden67,nakamura00}, or the close random stellar encounters (before the onset
of gravothermal oscillations, see \citealt{BT}, sect.~8.4, and references therein).

Nevertheless, the isothermal sphere has infinite mass and, moreover,
the spatial extent of a real system cannot be infinite, also because of
the presence of the external galactic field, thus some kind of
cutoff for high-energy stars at a given tidal radius $r_\mathrm{t}$ must be included.
Among the various possibilities, the `lowered' Maxwellian proposed by
\citet{michie63}, \citet{boden63} and \citet{king66}, on the basis of Fokker-Planck
calculations, has been the most successful one, both because of easy of
computation and implementation and because of the rather good agreement
with observations \citep{king81}.
Then, after the first results achieved with single component (i.e. equal
star mass) models \citep[e.g.,][]{illing76,illing77}, it was clear that to reproduce the
observed surface brightness profiles it is often necessary to include stellar
objects with different mass, i.e. multi-component DFs.

A general form for a $n$-component isotropic DF can be given as a linear
combination of single lowered Maxwellians \citep{dacosta,gunn}:
\beq
f(r,v)=\sum_{k=1}^n C_k f_k(r,v), \label{DF}
\eeq
with
\beq
f_k(r,v)=\left\{
\begin{array}{ll}
e^{-E/\sigma^2_k}-1,&
\textrm{ if $E<0$,}\\
0, & \textrm{ if $E\geq 0$,}
\end{array}\right.
\label{DFk}
\eeq
being $E=v^2/2+\Psi(r)$ the energy per unit mass and $\Psi(r)$ the mean
total potential
defined so to have $\Psi(r_\mathrm{t})=0$. The $C_k$ are
normalization constants related with the global mass function, and the
parameters $\sigma_k$ are connected to the velocity
dispersion and depend on the stellar mass $m_1,m_2,\dots,m_n$ of the components (and,
as we will see, on their concentration).

In these models, the inclusion of the star mass as a further degree of freedom
makes that another dynamical process can be taken into account: the energy
equipartition that is expected to occur in the denser regions via
2-body collisions between stars with different masses \citep{spitzer69}.
An important consequence of this phenomenon,
that has been found on various globular clusters by means of
{\it HST\/} (\citealt*{king95}; \citealt{sosin97}; \citealt*{albrow02}),
VLT \citep{andreuzzi01} and SDSS observations \citep{koch04},
is the `mass segregation', i.e. a `drift' of low-mass stars towards
orbits located mainly in the cluster outskirts while heavier stars
tend to concentrate in the inner region. For this reason, these
multi-mass models have been used also in $N$-body simulations
of clusters evolution to generate mass-segregated initial conditions
(\citealt{combes}; \citealt*{nostro}).

When accurate enough kinematical data became available, the so-called
King-Michie (KM) models, in which the DF of Eq.~(\ref{DF}) includes velocity
anisotropy through the dependence of $C_k$ on angular moment, have been
employed in order to reproduce the steep decrease
in the projected velocity dispersion observed in many clusters \citep{gunn}.
Indeed, some degree of anisotropy in the cluster outer regions should
be present as a reminiscence of the initial, quasi-radial, collapse
phase of cluster formation \citep{spitzer87}. This is also an important
issue from a stellar dynamics point of view, because it is still unclear up to what
extent this anisotropy can `survive' the isotropising effect of the tidal
field \citep{taka00,baum03}.

The most serious disadvantage of this `model-building' approach
is that many, and more and more, degrees of freedom have to be included
in order to improve the fit with the observed profiles
\citep*[see the discussion in][]{merritt97}
that were also proved to poorly constraint all the various parameters
\citep{dejonghe}.
For this reason, the alternative `non-parametric' approach was
proposed \citep{merritt93a,merritt93b} in which the relevant globular
clusters features can be directly extracted by the observed profiles
without making any assumptions on the form of the underlying
DF\footnote{Actually, some assumption must be made
in order to infer the $M/L$ behaviour and the mass function form \citep{geb95}.}.
Nevertheless, such an approach gives {\it no\/} direct indications on
what are the actual dynamical mechanisms that determine the
cluster evolution.
For this reason, in our opinion, the parametric technique can still provide
a valuable help in the comprehension of globular clusters dynamics,
especially when used in conjunction with the non-parametric approach
in order to test the validity of the various physical assumptions
made in writing the form of the given DF.
For instance, it still remains an intriguing problem that of understanding
why the form of the giants' DF found by \citet{geb95} in their non-parametric
study of the two non-collapsed clusters NGC 362 and NGC 3201, is clearly
non-Maxwellian in the central region.

By the way, it is crucial that all the dynamical mechanisms
taking into account in the parametric method are correctly and
coherently `implemented'.
Nevertheless, in all the KM multi-mass models
employed so far \citep[e.g.,][]{dacosta,gunn,pryor91,fischer93,cote95,meylan95} the
energy equipartition is ensured just by setting the square of the velocity
parameter of a star
component as $\sigma_k^2\propto 1/m_k$, which is a prescription that is
only asymptotically correct in the limit of isothermal DF.

Various authors \citep[e.g.,][]{merritt81,kondratev,heggiemeylan} have already
pointed out that such an `isothermal approximation' does not strictly
imply the central equipartition of kinetic energy between stars of
different mass. Moreover, \citet{merritt81} drew interesting conclusions on
the `equipartition instability' that he re-considered on models with
exact central equipartition. Nevertheless, one may still think that the isothermal
approximation leads, after all, to a central energy equipartition that is,
to practical purposes, always acceptable.
To this regard, following \citet{merritt81}, we construct
multi-component models in which exact equipartition of kinetic
energy is enforced at the centre (Sect.~\ref{exact}) and,
in Sect.~\ref{consequence}, compare
their properties and fitting parameters with those of models
constructed in the standard way.
Discussions and conclusions can then be found
in Sect.~\ref{concl}.

\section{The inclusion of central energy equipartition}
\label{exact}
In Eq.~(\ref{DFk}), the non-isothermal nature of $f_k$, due to
the presence of the term $-1$, makes each component velocity dispersion
to depend on the position $r$, such that {\it global\/} equipartition
is, in fact, impossible.
By the way, a global equipartition would be unrealistic for globular clusters since
in their outer regions relaxation time is supposed to be longer than their age.
Nevertheless, such an equipartition is expected to occur at the centre
($r=0$) where the relaxation time is shortest \citep{spitzer71,gunn}. This local equipartition
can be formally ensured by imposing the relations:
\beq
m_k\media{v_k^2}=m_1\media{v_1^2}, \label{equip}
\eeq
where the first mass class is chosen as the `reference' component, without the
loss of generality, and $v_k^2$ denotes the modulus square of the
velocity of the stars in the $k$-th component at $r=0$, i.e.
\beq
\media{v_k^2}=\int_0^w f_k(0,v)v^4dv
\left/\int_0^w f_k(0,v)v^2dv\right., \label{v_i}
\eeq
with $w=\sqrt{2|\Psi_0|}$ the escape velocity,
being $\Psi_0=\Psi(0)$ the central potential.

\begin{figure}
\includegraphics[width=8.5cm]{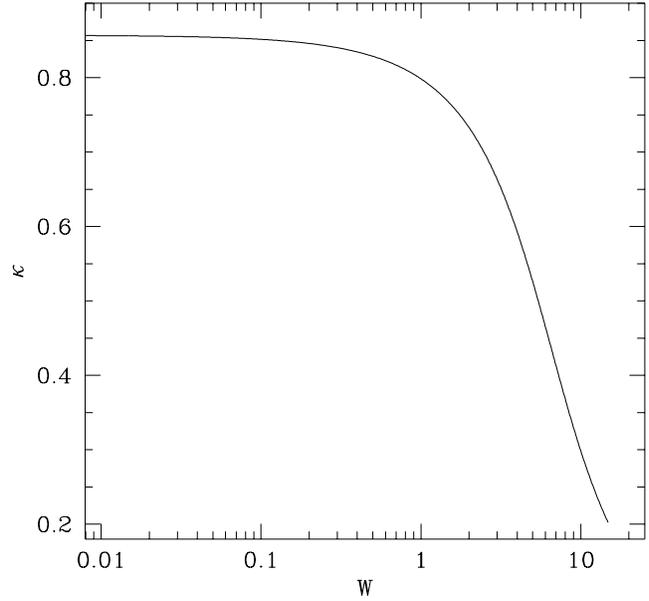}
\caption{Plot of the function $\kappa(W)$ defined in Eq.(\ref{kappa}). Note
that $\kappa(W)\to 6/7$ for $W\to 0^+$.
\label{fig_kappa}}
\end{figure}

In the following, $W_k\equiv|\Psi_0|/\sigma_k^2$ denotes the
dimensionless central potential and, moreover,
when referring to a quantity of the generic $k$-th component, for simplicity the
index $k$ is omitted (if unambiguous).\\
It can be shown that:
\[
w^{-3}\int_0^w(e^{-W}e^{-v^2/2\sigma^2}-1)v^2dv
\]
\beq
\textrm{\ \ }=\frac{\sqrt{\pi}}{4}
e^W W^{-3/2}\erf(\surd W)  - \frac{1}{2W} - \frac{1}{3}, \label{eq1}
\eeq
\[
w^{-5}\int_0^w(e^{-W}e^{-v^2/2\sigma^2}-1)v^4dv \nonumber
\]
\beq
\textrm{\ \ }=\frac{3\sqrt{\pi}}{8}
e^W W^{-5/2}\erf(\surd{W})  - \frac{3}{4W^2} - \frac{1}{2W} - \frac{1}{5},\label{eq2}
\eeq
with $\erf(x)\equiv(2/\sqrt{\pi})\int_0^x e^{-t^2}dt$ the error function.
Using the relations (\ref{eq1}) and (\ref{eq2}), Eq.(\ref{v_i}) can be rewritten,
after some terms rearrangement, as
\beq
\media{v^2}=\sigma^2W\kappa(W), \label{vsq}
\eeq
where one defines
\[
\kappa(W)
\]
\[ \textrm{\ \ }\equiv\left[\frac{3\sqrt{\pi}}{4} e^W\erf(\surd W)-
\frac{3}{2}W^{1/2} - W^{3/2} - \frac{2}{5}W^{5/2} \right]
\]
\beq
\textrm{\ \ }\times\left[\frac{\sqrt{\pi}}{4} W e^W\erf(\surd W) -
\frac{1}{2}W^{3/2} - \frac{1}{3}W^{5/2} \right]^{-1}. \label{kappa}
\eeq
Hence the constraint (\ref{equip}), imposes that
$m\sigma^2W\kappa(W)=m_1\sigma_1^2W_1\kappa(W_1)$.
Therefore, being $\sigma_1^2W_1=\sigma^2W=|\Psi_0|$, the central equipartition
leads to the following relation for $W$:
\beq
m\kappa(W)=m_1\kappa(W_1). \label{equip0}
\eeq
Given $W_1$ and $m/m_1$, the quantity $W$ can be found for any component
provided the knowledge of the inverse function of $\kappa$, which exists
for $W>0$ and $0<\kappa < 6/7$ -- being $\kappa(W)$ a monotonically
decreasing continuous function (Fig.~\ref{fig_kappa}) -- and which can be
easily computed by numerical means (e.g. using the bisection algorithm). Thus,
\beq
W=\kappa^{-1}\left(\frac{m_1}{m}\kappa(W_1)\right). \label{equip2}
\eeq
Then, once given $\sigma_1$, the parameter $\sigma$ for any other component
is determined by the relation:
\beq
\sigma^2=\sigma_1^2\frac{W_1}{W}. \label{sigma_i}
\eeq
Note that, since $\kappa < 6/7$,
Eq.~(\ref{equip0}) implies the existence of a {\it lower
limit\/} for the stellar mass in the components, in fact
one has that
\beq
m> \frac{7}{6}\kappa(W_1)m_1.\label{lowlimit}
\eeq
This is just a direct consequence of the obvious condition
$\media{v^2}<w^2$ and it means that, if exact equipartition is
achieved in the {\it central\/} region of a real cluster, objects with a mass
lower than the limit imposed by condition (\ref{lowlimit}) acquire enough energy
to escape from the system. Hence, they can be present there only
if they are bound in multiple systems
formed before 2-body relaxation takes place (eg., light remnants in
primordial binaries).
In this respect, note that for a component with such light objects, it
might be appropriate to multiply the $f_k$ of Eq.(\ref{DFk}) by a
factor that depopulates the phase-space region with low angular
momentum $L$, so as to make its inclusion in the model consistent with
the equipartition. A suitable factor could be, e.g.,
\beq
1-\exp\left(-L^2/2r^2_\mathrm{av}\sigma_k^2\right),
\eeq
being the `avoidance radius', $r_\mathrm{av}$, a further free parameter.
Notice, finally, that with the typical values $m_1=0.7 \msol$ and $W_1=8$,
Eq.~(\ref{lowlimit}) yields $m> 0.3 \msol$.

\subsection{The isothermal approximation}\label{isoapp}
As known, the central dimensionless potential $W$ determines the
`form', the shape, of the density distribution given by the $f_k$ of
Eq.~(\ref{DFk}), being the other parameters (e.g. $\sigma$ and $\rho(0)$)
just scaling factors.
In particular, $W$ is directly related to the King {\it concentration\/}
coefficient $c=\log(r_\mathrm{t}/r_\mathrm{K})$, with $r_\mathrm{K}\equiv 3\sigma(4\pi G\rho_0)^{-1/2}$
the King radius; they satisfy
the approximate linear relation $c\simeq 0.06+0.2W$ \citep{king66, BT}.
Then, when $W\gg 1$ it is clear that the presence of the tidal cut will
have a negligible role, in fact one can
say that $f_k\simeq\exp(-E/\sigma_k^2)$ at the centre and for
small velocities ($v^2\ll W\sigma^2$), hence the $f_k$ approaches an
isothermal DF whose velocity dispersion is $\media{v^2}= 3\sigma^2$.
Thus, the `isothermal approximation' (hereafter IA)
yields, from Eq.~(\ref{equip}), the commonly adopted equipartition
condition
\beq
\sigma^2\simeq \sigma_1^2\frac{m_1}{m}, \label{equip_iso}
\eeq
which allows to determine $\sigma$ for any component and, in turn, to evaluate
its central dimensionless potential as
$W_\mathrm{IA}= W_1\sigma_1^2/\sigma^2=mW_1/m_1$.

Note that under the IA, Eq.(\ref{kappa}) correctly leads to the same condition
since the $e^W$ term dominates for $W\gg 1$, giving $\kappa(W)\simeq 3/W$,
hence Eq. (\ref{equip0}), for $W\gg 1$ and $W_1\gg 1$,
gives just $W\simeq mW_1/m_1$.

\begin{figure}
\includegraphics[width=8.5cm]{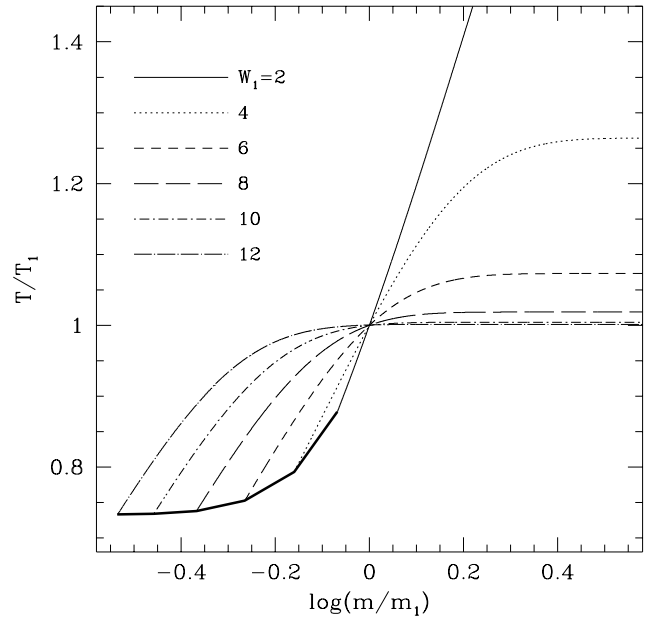}
\caption{Temperature ratio at the cluster centre
between a generic component with mass $m$ and a `reference' one
having mass $m_1$ and $W_1$ as labelled, when the energy
equipartition is imposed using the `isothermal approximation'.
The thick line marks the low-mass cutoff given by (\ref{lowlimit}).
\label{fig_tratio}}
\end{figure}

It is now worth checking and quantifying more precisely the goodness of the
IA: let us evaluate the actual `temperature' ratio $T/T_1$
between a generic component and the reference one at the centre of the cluster,
and compare it with the unitary value actually demanded by the equipartition.
From Eq.~(\ref{vsq}) we have that
\beq
\frac{T}{T_1}=\frac{m\media{v^2}}{m_1\media{v_1^2}}
=\frac{m\kappa(W_\mathrm{ IA})}{m_1\kappa(W_1)}
=\frac{m\kappa(mW_1/m_1)}{m_1\kappa(W_1)}.
\label{tsut1}
\eeq
In Fig.\ref{fig_tratio} this ratio is plotted for
various values of $W_1$ as a function of the mass ratio $m/m_1$. As expected,
the isothermal approximation works better, i.e. $T/T_1$ is closer to the unit,
for higher $W_1$ and for $m>m_1$, because in these regimes the DF of both
components are closer to the isothermal distribution.
However, although one can affirm that the dynamical condition of central
`thermal equilibrium' is approximately valid within a limited range of masses
and for sufficiently concentrated models, nevertheless the model
structure is rather sensitive to the temperature ratio.
Let us consider the quantity
\beq
\frac{W_\mathrm{IA}}{W}=\frac{mW_1}{m_1W}, \label{wiso}
\eeq
where $W$ is given by Eq.~(\ref{equip2}).
This ratio quantifies the change affecting the fundamental structural
parameter of the clusters profile, when the IA is adopted instead of
the exact energy equipartition.

\begin{figure}
\includegraphics[width=8.5cm]{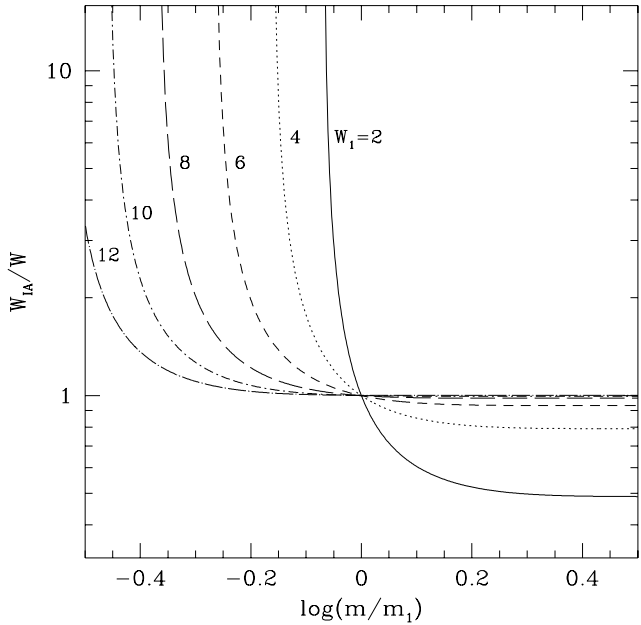}
\caption{Ratio between the dimensionless central potential
evaluated under the IA of a component with mass $m$, and
that turned out by the correct energy equipartition
with a `reference' component having mass $m_1$ and various
$W_1$. The vertical asymptotes correspond to the
$m$ lower limit (\ref{lowlimit}).\label{fig_epsilon}}
\end{figure}

As shown by Fig.\ref{fig_epsilon}, one can see that
$W_\mathrm{IA}$ can be {\it very\/} different from $W$, especially for low
concentration models, e.g. for non-collapsed globular clusters ($W_1\la 10$).
In particular, given a pair of components, the use of the IA implies a
{\it higher\/}
concentration for the component with lighter stars, and a {\it lower\/}
concentration for the heavier stars.
In other words, with respect to the correct energy equipartition assumption,
the IA leads to a {\it reduced\/} mass segregation.

Then, as a consequence on the parametric fitting of globular clusters
profiles, one expects that, once adjusted the distribution of the brightest stars component
(giants and turn-off stars) to fit the observed luminosity profile,
the IA gives: (i) a steeper and too much concentrated mass density
profile in the outer regions, being this determined by the lighter (dwarf main sequence)
stars mostly; (ii) a flatter density profile for the massive remnants in the core
region; (iii) a steeper radial velocity dispersion profile for the giants, as a
consequence of the steeper decreasing of the potential.
Point (i) implies then a lower total cluster mass, and so a lower
mass-to-luminosity ratio, in respect to the exact energy equipartition.

Notice that all this discussion remains valid also in the case of
KM models where the presence of velocity anisotropy is considered.
Indeed, in these models, the coefficients $C_k$ in Eq.(\ref{DF}) are equal to
$\alpha_k \exp(-r^2v^2_\perp /2r^2_\mathrm{a}\sigma^2_k)$, where $r_\mathrm{a}$ is
the length-scale above which anisotropy is important and $v_\perp$ is the further
state variable corresponding to the modulus of the tangential velocity
(see Appendix \ref{construction}).
Thus, for $r=0$ there
is no anisotropy and $C_k=\alpha_k$ is independent of $v^2$.

\section{Consequences on the observable profiles and relevant parameters}
\label{consequence}
\begin{figure}
\includegraphics[width=8.5cm]{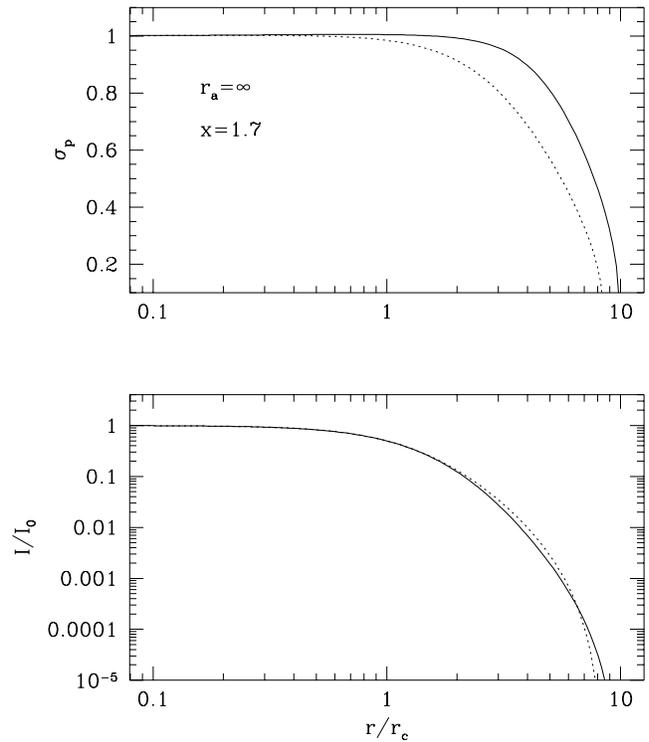}
\caption{Upper panel: projected velocity profile for the
luminous component of the isotropic 3-component model described
in the text. Lower panel: projected surface brightness profile.
Solid line: profiles obtained in the case of exact energy
equipartition among the components; dotted line: profiles
obtained under the IA.
\label{prova1}}
\end{figure}

To understand better the effects of the IA on the models morphogical and kinematical
features, a simple case of an  isotropic ($r_\mathrm{a}=\infty$) 3-component model
is considered. Though representing a crude simplification in respect to real
clusters with a continuous mass spectrum, it is sufficient to provide significant
indications on the consequences of the approximation we are discussing.
The model is made up of:
(i) giant and turn-off stars ($m_1=0.75\msol$, total mass $M_1$), determining
the light profile;
(ii) main sequence dwarf stars and
light remnants ($m_2=0.26\msol$, $M_2/M_1=22$), considered to be dark, but
dominating the gravitational potential; (iii) heavy dark remnants
($m_3=1.2 \msol$, $M_3/M_1=0.28$, remnants of objects with mass in the
range $4$--$8 \msol$ ).
All is consistent with a mass function $dN\propto m^{-x}d\log m$ with $x=1.7$,
and with condition (\ref{lowlimit}).
The surface brightness profiles correspond to a concentration parameter
$c\simeq 1$.
In Fig.~\ref{prova1} the surface brightness and the projected
line-of-sight velocity dispersion profiles given adopting the IA and the
exact equipartition, are shown for a comparison. The radius is given
in unit of the {\it core\/} radius, $r_\mathrm{c}$, at which the
luminosity drops to $1/2$ the central value.
The parameter $W_1$ has been adjusted in such a way to give similar
light profiles in both cases.

\begin{figure}
\includegraphics[width=8.5cm]{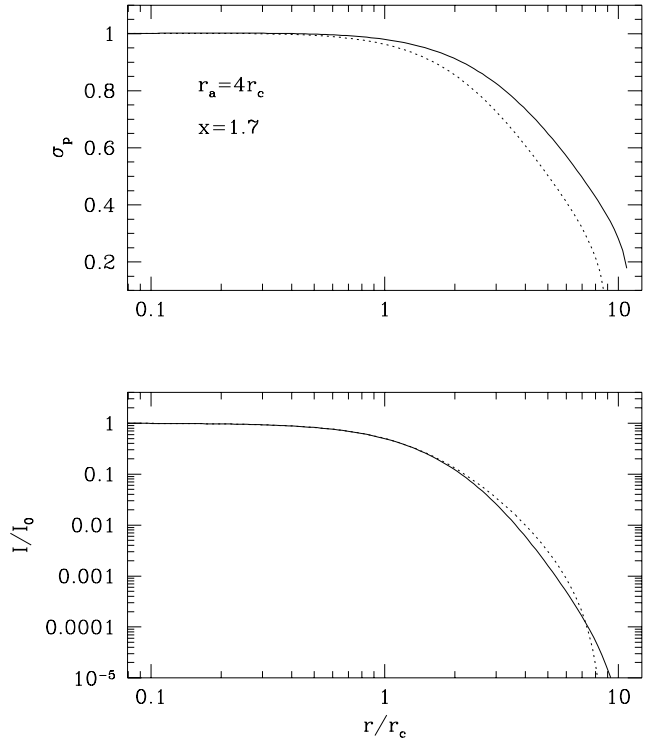}
\caption{The same as in Fig.~\ref{prova1} but with an anisotropy radius
$r_\mathrm{a}=4r_\mathrm{c}$ \label{prova1an}}
\end{figure}

\begin{figure}
\includegraphics[width=8.5cm]{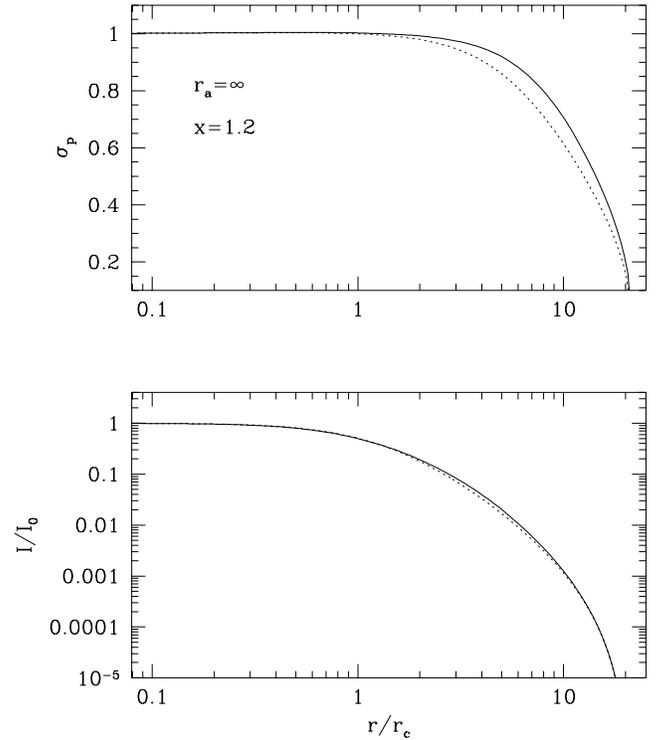}
\caption{The same as in Fig.~\ref{prova1} but with a flatter mass
function ($x=1.2$). \label{prova2}}
\end{figure}

Thus, confirming the points discussed at the end of Sect.~\ref{isoapp},
the IA gives a $\sigma_\mathrm{p}$ profile appreciably steeper than in
the correct case.
The cluster envelope is also too concentrated ($W_2=2.3$ instead of the
correct $W_2=8.7\times 10^{-2}$) and the massive remnants
have a much flatter density profile in the core region ($W_3=11$
instead of $W_3=16$).
Though to a less extent, this occurs also in the presence of velocity
anisotropy, as shown in Fig.\ref{prova1an}.
Moreover, the difference is lower also when a less peaked mass function
is adopted, as indicated in Fig.~\ref{prova2} that refers to the same isotropic
model above-described, but with $x=1.2$, i.e. with:
$m_1=0.75\msol$; $m_2=0.28\msol$, $M_2/M_1=14$; $m_3=1.2 \msol$, $M_3/M_1=0.76$.
Of course, in this case the reduced discrepancy is due to the less
importance of the lightest component in determining the overall potential.

\begin{figure}
\includegraphics[width=8.5cm]{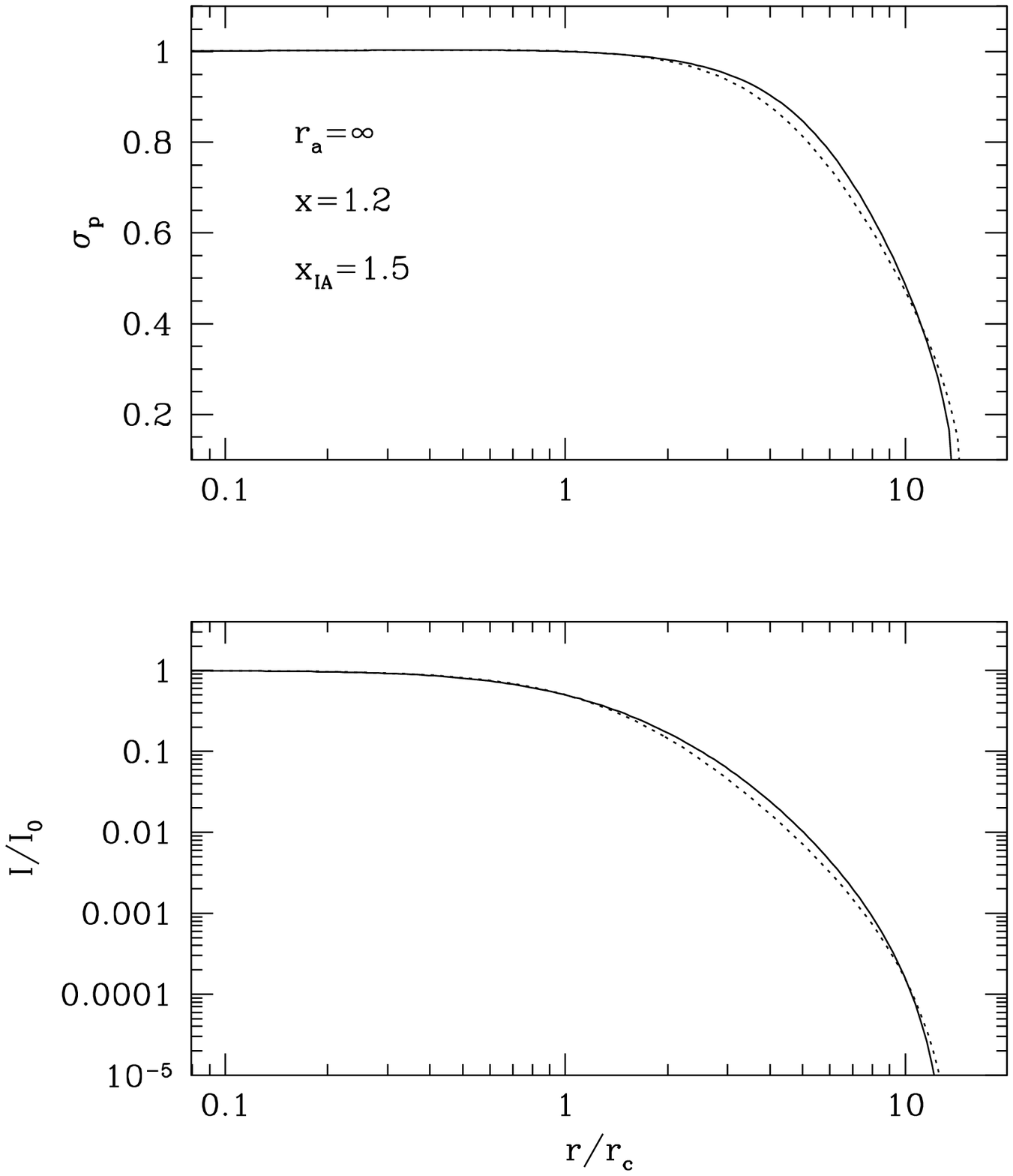}
\caption{Projected velocity dispersion (upper panel) and surface brightness
(lower panel) profiles for the luminous component of the isotropic
3-component model made under the IA (dotted line) and with the correct energy
equipartition (solid line). They give similar curves with different
fitting parameters (see Table~\ref{tab1}).
\label{prova_coinc}}
\end{figure}

Now, to quantify the change affecting all the relevant parameters
because of the use of the IA, these have to be compared with the
fitting parameters that, under the correct energy equipartition, give
the same projected surface brightness {\it and\/} velocity
dispersion profiles. In Fig.~\ref{prova_coinc}
the profiles corresponding to a typical non-collapsed
cluster ($c=1.2$ in this case), are plotted. The resulting
parameters are written in Table~\ref{tab1}. Besides the difference
in the lightest component concentration, one can immediately see
the discrepancy in the total mass (and so in the mass-to-light ratio)
and in the mass function index. In this example, the IA leads
to a $1.4$ times lower total mass and to a steeper
mass function, $x=1.5$ instead of $1.2$, giving an amount of mass
in heavy remnants which is $3.3$ times lower.
\begin{table*}
 \centering
 \begin{minipage}{140mm}
  \caption{Fitting and derived parameters for the 3-component isotropic model
whose profiles are plotted in Fig.~\ref{prova_coinc}. The reported parameters
are: the mass function logarithmic slope ($x$) and the system total mass ($M$, in
arbitrary units); then, for the components: the
total mass ($M_k$, in arbitrary units), the central dimensionless
potential ($W_k$), the velocity parameter ($\sigma_k$, with $\sigma_1=1$)
and the temperature ratio. In both cases we have
$m_1=0.75$, $m_2=0.28$ and $m_3=1.2 \msol$.
\label{tab1}}
  \begin{tabular}{@{}p{2.cm}cccccccccccc@{}}
  \hline
energy equipartition &
$x$&
$M$&
$M_1$&
$M_2$&
$M_3$&
$W_1$ &
$W_2$ &
$W_3$ &
$\sigma_2$ &
$\sigma_3$ &
$T_2/T_1$ &
$T_3/T_1$\\
\hline
under IA&
$1.5$&
$11$&
$0.58$&
$10$&
$0.24$&
$9.1$&
$3.3$&
$15$&
$1.7$&
$0.79$&
$0.75$&
$\sim 1$\\
exact&
$1.2$&
$16$&
$1.0$&
$14$&
$0.79$&
$9.2$&
$5.9\times 10^{-2}$&
$15$&
$12$&
$0.79$&
$1$&
$1$\\
\hline
\end{tabular}
\end{minipage}
\end{table*}
See Appendix \ref{construction} for a detailed description of
the procedure followed to construct the KM models.

\section{Conclusion and discussion}
\label{concl}
In the multi-mass King-Michie models usually employed for the study of morphology and
kinematics of globular clusters, the energy equipartition at the cluster centre,
between stars of different masses, is imposed by setting the square of the
velocity parameter of the $k$-th mass class, $\sigma^2_k$ in the DF of Eq.~(\ref{DF}),
inversely proportional to the mass, $m_k$, of that component.
As \citet{merritt81} and \citet{kondratev} emphasized, this is correct only
in the limit of isothermal distributions (i.e.
with central dimensionless potential $W\to \infty$). This `isothermal
approximation' (IA) has been checked and the consequences studied in this paper.
To this purpose, a general procedure to construct $n$-component models with
exact energy equipartition at the centre is used and described in detail.

From the point of view of the `thermal equilibrium', it is found that the
`temperatures' of a pair of components can differ to a non-negligible
amount for low concentration distributions. For instance,
taking a giant and turn-off stars component with mass $m_1\sim 0.8\msol$
having a concentration $c\sim 1$, along with a typical dwarf main-sequence
star component with $m_2\sim 0.2$, one has that, under the IA, this latter
component has a temperature ($m\media{v^2}/2$) which is actually a factor
$0.75$ lower than that of the giants.

To investigate the consequences on the observable profiles,
a simple 3-component isotropic model was generated resembling roughly that of a
typical non-collapsed cluster. Aside from the two above-mentioned components,
heavy remnants with $m_3=1.2\msol$ were also included, while the components
total mass were varied according to different mass function logarithmic
index $x$. Once adjusted the giant profile to match the same light
distribution, it is found that (with $x=1.7$) the cluster envelope is
much less concentrated in the exact equipartition case ($W=8.7\times 10^{-2}$)
than under the IA ($W=2.3$).
Moreover, the IA gives a flatter density profile for the massive remnants
in the core region ($W=11$ instead of $W=16$).
As regards the velocity dispersion profile, the approximation leads to a
steeper profile for the giants, as a consequence of the steeper decreasing
of the potential.
However, it is found that both the presence of anisotropy and lower mass
function slopes reduce the variations in respect with the correct case.

Then, in order to evaluate the change in the most relevant parameters,
these were fixed so as to give the same projected surface brightness
profile and the same line-of-sight velocity dispersion curve,
with models constructed both with the correct
energy equipartition and employing the IA. It is found that the inferred
total mass (and, correspondingly, the global mass-to-light
ratio) results $1.4$ times lower than the correct value, with a steeper
mass function that leads to a $3.3$ times
smaller amount of mass in heavy dark remnants.

By the results of this work it can be stated that the IA leads to
multi-mass parametric fit of non-collapsed globular clusters profiles,
corresponding to models {\it not\/} consistent with the assumption
of energy equipartition at the centre. Consequently, the estimate of
many important parameters (such as: total mass, mass-to-light ratio, mass
function slope, heavy remnant abundances, etc.) is appreciably affected
by the non-unitary ratio among central components temperatures.
Moreover, in the multi-mass models used so far, condition (\ref{lowlimit}) is
usually ignored. Indeed, it is assumed the presence of centrally concentrated
light objects (light remnants or dwarf main sequence stars, with mass as low
as $0.1 \msol$) that, on the contrary, should not be present in the central
region, since they acquire there sufficient kinetic energy to leave the system.

Thus, though the models with IA can reproduce rather well the
observable profiles, they are inconsistent with one of the dynamical
processes (energy equipartition) they rely on. This is
incoherent with the underlying `philosophy' of the parametric
approach. By the way, if some dynamical constraints were relaxed
without any theoretical justification, then non-parametric techniques
would definitively be more appropriate.

Whether or not the energy equipartition takes actually place
at the cluster centre is an entirely different problem (to this regard,
our opinion contrasts to what claimed in \citealt{kondratev}),
framed into a still controversial subject.
In this respect, it is worth noting that \citet{pryor91} deduced,
thanks to a parametric study of low-concentration clusters, a
relatively high and sharp cutoff in the mass function, at $0.3$--$0.4 \msol$,
for NGC 5466 and M55, which is in good agreement with condition (\ref{lowlimit}).
Furthermore, the \citet*{romaniello95} and \citet{demarchi95} {\it HST\/}
deep photometric studies, concerning with the region around the half-mass radius
of various clusters, indicates a general remarkable flattening of the mass
function below $0.3 \msol$. This was also observed, for
higher masses, in the core of 47 Tuc \citep*{paresce95}. Then, energy equipartition is, at least, not contradicted by these
observational evidences about the lacking of low-mass stars in the central
regions of clusters.

From a theoretical point of view, besides the efficiency of 2-body
collisions, also the relative importance that the different components
have in determining the total gravitational potential influences much
the equal distribution of kinetic energy among them and, as firstly
envisaged by \cite{spitzer69}, it could be impossible, under
certain conditions, to be attained. In a few words, this happens when
a much more massive stellar species (with mass $m_2$) has a
negligible influence on the overall potential determined by a much
lighter one ($m_1\ll m_2$, but $M_1\gg M_2$), so that
the heavier stars get more and more concentrated (while the others expand),
becoming a self-gravitating system with a negative heat capacity. Thus,
the more kinetic energy it loses, the higher will be its temperature
departing faster and faster and more and more away from the energy
equipartition with the lighter component, leading to the so-called
`equipartition instability'.
Though \citet{inag85} showed that equipartition
never occurs in their Fokker-Plank models, \citet{spitzer71} and the
more recent Monte Carlo simulations by \citet*{watters00} have basically
confirmed the original Spitzer's criterion for a 2-component model.
In this respect, one must note that neither the relation
\beq
\left(\frac{M_2}{M_1}\right)\left(\frac{m_2}{m_1}\right)^{2.4}\leq 0.32
\label{spitz}
\eeq
suggested by \citet{watters00} for equipartition to occur, nor the
Spitzer's one, are satisfied for
the 2nd and the 3rd components of our example models.
We must remind, however, the \citet{merritt81} conclusion, drawn by
re-considering the Spitzer's argument on models with exact central
equipartition, affirming that this latter can be reached even
if Spitzer's condition is violated.
Perhaps, more convincing indications could be provided, in the near
future,
by direct and collisional $N$-body simulations with a sufficiently
large $N$ and, more important, with a realistic mass function
\citep[see the introductory discussion in][]{inag85}.


To conclude, it can be stated that in the parametric study of
low-concentration clusters, it appears appropriate to
take particular care to the inclusion of the energy equipartition
process in KM multi-mass models,
since the structure of each star component and, consequently, all the
relevant parameters, are significantly influenced by the presence of
thermal equilibrium at the centre. Furthermore, the coherence of the
parametric approach must be preserved to make significant comparison
with the results of non-parametric (or other) techniques.

\section*{Acknowledgments}
The author wants to thank Prof. D.C. Heggie
for his crucial and helpful comments and Dr. P. Di Matteo for the stimulating
discussions about multi-mass models.
The author is also grateful to Mr. D. Miocchi for his help and support
and to the anonymous referee for his valuable suggestions.

\onecolumn
\appendix
\section{The construction of the King-Michie model}
\label{construction}
Let us consider the anisotropic King-Michie DF
\beq
f(r,v,rv_\perp)=\sum_{k=1}^n f^{\mathrm a}_k(r,v,rv_\perp)=\sum_{k=1}^n \alpha_k
\exp(-r^2v^2_\perp /2r^2_\mathrm{a}\sigma_k^2) f_k(r,v),
\label{DFa}
\eeq
where $f_k$ is that of Eq.(\ref{DFk}) and $v_\perp=v\sin\eta$, with $\eta$
the angle between the velocity and the radial direction.
To construct a self-consistent model it is necessary that the potential
$\Psi(r)$ in the DF is that generated, according to the
Poisson's equation
\beq
\frac{d^2\Psi}{dr^2}+\frac{2}{r}\frac{d\Psi}{dr}=4\pi G \rho (r),
\label{poisson}
\eeq
 by the mass density
\beq
\rho(r)=\sum_{k=1}^n\rho_k=2\pi\sum_{k=1}^nm_k\int_0^w
\left[\int_0^\pi f_k^\mathrm{a}(v^2/2+\Psi(r),rv\sin\eta)
\sin\eta d\eta\right]v^2dv \label{dens}
\eeq
determinated,
in turn, by the DF itself.
As done in \citet{gunn}, the problem is solved through
the numerical integration of Eq.~(\ref{poisson}), with the initial
conditions $\Psi(0)=-W_1\sigma_1$, $\Psi'(0)=0$, employing
a 4-th order Runge-Kutta method.
However, a completely analytical expression was used for
the density, without the needing of numerical quadratures.
Indeed, it can be shown that, for the $k$-th component, the integral
in Eq.~(\ref{dens}) gives:
\beq
\rho_k(\Psi,r)=\alpha_k\frac{m_k(\sqrt{2\pi}\sigma_k)^3}{\tilde r}
\left[\left(\frac{1}{\tilde r^2}-
\frac{1}{1+\tilde r^2}\right)
\exp\left(-\tilde r^2{\cal W}_k\right)
\erfi\left(\tilde r\surd {\cal W}_k\right)+
\frac{\tilde r}{1+\tilde r^2}
\exp\left({\cal W}_k\right)
\erf\left(\surd{\cal W}_k\right)-
\frac{2}{\sqrt{\pi}}\frac{\surd{\cal W}_k}{\tilde r}
\right], \label{dens2}
\eeq
where ${\cal W}_k\equiv -\Psi/\sigma_k^2$, $\tilde r\equiv r/r_\mathrm{a}$
and $\erfi(x)\equiv (2/\sqrt{\pi})\int_0^x e^{t^2}dt$ is the
imaginary error function
that can be computed through truncated power series expansion, so as
to have:
\beq
\sqrt{\pi}e^{-x^2}\erfi(x)\simeq
\left\{
\begin{array}{ll}
2e^{-x^2}\sum_{j=0}^p [j!(2j+1)]^{-1}x^{2j+1},&
\textrm{ if $x\leq 2$,}\\
\sum_{j=0}^p (2j-1)!!2^{-j}x^{-(2j+1)}, & \textrm{ if
$x> 2$}
\end{array}\right.
\eeq
that, with $p=4$, gives a truncation error $<1$ per cent for $x\geq 0$.
On the other hand, being $e^{-x^2}\erfi(x)\sim (2/\sqrt{\pi})(x-2x^3/3)$
for $x\to 0$, it is easy to verify that, for $\tilde r \ll 1$,
Eq.~(\ref{dens2}) tends to the correct expression for the density in
the isotropic case \citep[][eq.~4-131]{BT}. Moreover,
the asymptotic behaviour
$\rho_k(\Psi,r)\simeq 8\pi\alpha_k m_k(\sqrt{2}\sigma_k)^3{\cal W}^{\, 5/2}_k/15$,
for ${\cal W}_k\ll 1$, can be used to avoid round-off error in numerical evaluations.

The input needed to construct a complete self-consistent multi-mass model
with anisotropic velocity is, then:
\begin{description}
\item the components star mass $\{ m_1,m_2,\dots,m_n\}$;
\item the components total mass $\{ M_1,M_2,\dots,M_n\}$ and the mass-to-light ratio
$\{ \Upsilon_1,\Upsilon_2,\dots,\Upsilon_n\}$;
\item the velocity parameter of the `reference' component $\sigma_1$;
\item the central density of the ref. component $\rho_{0,1}$;
\item the dimensionless central potential of the ref. component $W_1$
(which gives $\Psi_0=-W_1\sigma_1$);
\item the anisotropy radius $r_\mathrm{a}$ (assumed to be the same for all components).
\end{description}
Then, in order to impose energy equipartition at the centre, $\sigma_k$ for
the other components has to be fixed by the relation
$\sigma_k=\sigma_1W_1/W$, being $W$
given by Eq.(\ref{equip2})
with $m=m_k$. On the contrary, if one wants to test the isothermal approximation
then $\sigma_k=m_1\sigma_1/m_k$.

The first normalization constant $\alpha_1$ is readily fixed by
the requirement $\rho_1(\Psi_0,0)=\rho_{0,1}$, while the
others are constraint by the given total component masses.
Since $M_k$ depends, through $\rho_k$, on the
overall potential which, in turn, depends on all the normalization
constants, these contraints correspond to a system of $n-1$ integral
equations for the set of $\alpha_{k>1}$. As usual, this
system is solved iteratively, starting from an initial guessed set of
$\alpha_k$, e.g. $\alpha_k\equiv \alpha_1M_k/M_1$,
halting the process when the difference between two subsequent sets
of $\alpha_k$ is within a given tolerance.

At the end, given $\rho(r)$ and the potential $\Psi(r)$, all the relevant
quantities can be evaluated using also Eq.~(\ref{DFa}). Thus,
the projected line-of-sight velocity dispersion of the $k$-th component
is given by \citep{richstone84}:
\beq
\sigma_{\mathrm{p},k}^2(R)=\frac{4\pi}{S_k(R)}\int_R^{r_\mathrm{t}}\frac{rdr}
{(r^2-R^2)^{1/2}}\int_0^1 dx
\left[ x^2\left( 2-\frac{3R^2}{r^2}
\right)+\frac{R^2}{r^2}\right]
\int_0^w m_kf_k^{\mathrm a}(r,v,rv_\perp)v^4dv
\eeq
where $x\equiv \cos\eta$, $v_\perp=v(1-x^2)^{1/2}$ and $S_k(R)$,
the projected surface density,
is
\beq
S_k(R)=2\int_R^{r_\mathrm{t}}\frac{\rho_k(r)r}
{(r^2-R^2)^{1/2}}dr.
\eeq
Of course, the projected surface brightness will be given
by $I(R)=\sum_{k=1}^n \Upsilon_k^{-1}S_k(R)$.

\bsp

\label{lastpage}

\end{document}